\documentclass[prb,amssymb,aps]{revtex4}
\usepackage{graphicx}
\usepackage{amsmath,amsthm,amssymb}

\begin{document}
\draft
\title{Two thermodynamic particularities of the dynamic glass transition in
liquids: Glarum-Levy defects and Fischer speckles $-$ cosmological
consequences}
\author{E. Donth}
\address{Institut f\"{u}r Physik, Universit\"{a}t Halle,\\
D-06120 Halle (Saale), Germany\\
E-mail: donth@physik.uni-halle.de}

\begin{abstract}
A new thermodynamics for liquids related to von Laue's approach (1917)
substitutes some particle priors of Gibb's rational thermodynamics. This \
allows the definition of a new dynamic entity ($G$ defect) whose diffusion
properties also claim a largest causal region ($F$ speckle). In the frame of
the hidden charge model \cite{this part} it is discussed, whether this new
thermodynamics can be applied to an initial liquid for cosmology, where the $%
G$ defects lead to the later galaxies and the $F$ speckles to a finite
expanding universe of diameter $R$. Far below a ''hadronic'' Compton wave
length $\lambda _{0}$ of order 1 fermi, $R\ll \lambda _{0}$, there is no
room left for too small filter elements, that would, however, be necessary
for a filter convergence to an isolated cold quantum mechanical point
particle. When the expansion of the universe comes to $\lambda _{0}$, i.e.
for $R\approx \lambda _{0}$, then many hadrons are created. The related
negative pressure gets large amounts and leads to an intense hadronic
cosmological inflation. $-$ The $G$ defects are formed by the shaping power
of Levy distribution (preponderant component, hierarchy, damping factor). A
relation between the number of galaxies, the tilt in the density
fluctuation, and the temperature amplitude of the CMB is obtained. For $R\gg
\lambda _{0}$ (much vacuum), the ''electromagnetic'' $M_{4}$ tangent objects
are large and flat. This allows a geometric interpretation of the ''stony''
dark energy as flatness on the ''golden'' side of the Einstein equation, $%
\Omega _{\Lambda }=1-\Omega _{M}$.


\ \ \ \ \ \ \ \ \ \ \ \ \ \ \ \ \ \ 2. power spectrum

\ \ \ \ \ \ \ \ \ \ \ \ \ \ \ \ \ \ 3. dark energy theory

\ \ \ \ \ \ \ \ \ \ \ \ \ \ \ \ \ \ 4. gravity

\end{abstract}

\maketitle

\newpage
\tableofcontents
\newpage

\section*{1. Introduction}

About 1960, a change in education and evaluation of general phenomenological
thermodynamics was observed. Before 1960, it was mainly based on arguments
by Carnot, Kelvin and Clausius. After 1960, it was mainly based on the Gibbs
distribution of statistical mechanics. This Gibbs rational thermodynamics
has the following priors for a system of interacting structureless particles:

\begin{enumerate}
\item The only general entities are the particles; the rational is obtained
from the thermodynamic limit: number of particles $N\rightarrow \infty $
under the condition of finite density ($N/V=$ given),

\item There is a large heat reservoir that feeds the temperature $T$ as an
equivalence-class index into the system,

\item The non-fluctuation of the reservoir temperature, $\delta T=0$, is
transferred into the definition of the general temperature (to justify the
sharp $k_{B}T$ denominator in the Boltzmann factor for energy in the Gibbs
distribution).
\end{enumerate}

These priors are tried and tested in systems where the spatial infinities,
sufficiently separated from the particles, are contained inside the system
itself, quasi as an internal heat reservoir. Examples are the infinite
lattice in a solid or an infinite vacuum of vacuum elements in a gas with isolated particles
imbedded. We ask, whether these priors can be applied to liquids, especially
to liquid dynamics, where such separated infinities are not obvious.\newline
\newline

The existence of two experimentally well-tried particularities could be
useful for cosmology by starting the universe from an initial liquid
(initial liquid hypothesis). Smaller ones (Glarum Levy defects for dynamic
heterogeneity of the dynamic glass transition, $G$ defect as a new entity)
could be the seeds for the later galaxies $-$ we use now the symbol $N$ for
their number in the universe $-$, and a larger one (Fischer speckle) could
be the largest causal region \ in the initial liquid, wherefrom the
increasing diameter $R(t)$ of the expanding universe results.\newline
\newline

In the dispersion zone of the dynamic glass transition of a classical
molecular liquid, the $G$ defects must dynamically be described by a Levy
distribution with a Levy exponent $\alpha $, $0<\alpha <1$ for relaxation
(frequency $\omega \lesssim 10^{11}$rad/s). This follows from the
representativeness theorem (section 3.1). The difference $1-\alpha >0$ is
called tilt and is later related to the tilt $1-n_{S}$ of the spectral index
$n_{S}$ in cosmic density fluctuations. The preponderant component of the
Levy sum for $\alpha <1$ corresponds to a Levy diffusion step of a molecule
through the cage door of its nearest neighbor particles. A relation between
the small $G$ defects and the large $F$ speckle is obtained via an
''analytic continuation'' of the diffusion steps. If the application to the
cosmological initial liquid is reliable, a relation results between the
number of galaxies $N$ in the universe of diameter $R$ and the tilt $1-n_{S}$%
. All this is worded by the phrase ''shaping power of Levy distribution''.%
\newline
\newline

The thermodynamic entity is the key concept of this second paper \cite%
{second part}. This entity is defined dynamically by a representative
containing one $G$ defect which is formally
labeled by one Boltzmann constant $k_{B}$ from an internal quantum
mechanical experiment of Nyquist (1928), cf. the Remark of section 3.1. The internal experiment is a
''self-experiment'' in the liquid $-$ without any reservoir and without any
conscious observer $-$ that is thermodynamically described by the
fluctuation dissipation theorem (FDT).\newline
\newline

This self-experiment may be understood as a common filter construction of
the hidden charge model \cite{this part}, common in relation to several
particles, or to quantums in a Nyquist transmission line. In the model, a
common filter \ construction from several eigensolutions must be allowed for
the construction of e.g. a baryon: from a 1-class lepton, a 2-class
prebaryon and a 3-class confinon eigensolution. The entity is, so to speak,
a common thermodynamic ''culminating event'' of the model.\newline
\newline

The term defect diffusion was introduced by Glarum \cite{Glarum}. An
experimental retardation function that could later be identified with the
characteristic function of the symmetric Levy distribution was detected by
Kohlrausch \cite{Kohlrausch} in 1847. Many experimental hints of the
connection of glass relaxation data with an underlying Levy distribution
where collected, e.g. by Ngai \cite{Ngai}. A theoretical review was e.g. by
Shlesinger \cite{Shlesinger}. The Fischer dispersion zone \cite{Fischer} and
his speckles \cite{Patkowski} are reviewed by himself. The Levy character of
the defects was further developed in the book \cite{Donth2001} containing
also the representativeness theorem. The FDT as an equation for
thermodynamic self-experiments was developed on the base of the Nyquist
model \cite{Nyquist}, first in 1982 \cite{Donth1982} and later sophisticated
in \cite{Donth2001}. Glass transition arguments were used in cosmology e.g.
in the paper of She \cite{She}. The hidden charge model is
discussed in part 1 \cite{this part}.

\section*{2. Thermodynamic entities. von Laue vs. Gibbs thermodynamics}

In classical liquids, local Glarum-Levy defects ($G$ defects) are necessary,
densily neighbored dynamic phenomena. The Levy property is a consequence of
the representativeness theorem (section 3.1), their locality (finite volume,
equation (\ref{Eq.2.2}) below) is a consequence of the local high-frequency
break throughs they are (''levy instability'' \cite{Donth2001}, otherwise a
general negative pressure would be obtained), and the dense arrangement is a
consequence of their mutual coexistence condition emerging from spatial
aspects of the individual Levy damping factors ($n^{-1/\alpha }$). The $G$
defects are connected with thermodynamic (= internal quantum mechanical)
self-experiments. $-$ On the other hand, the rational Gibbs thermodynamics
is not the consequence of such a self-experiment, if a large external heat
reservoir $-$ i.e. a macroscopic apparatus or observer $-$ is considered to
be necessary.

\subsection*{2.1 Temperature fluctuation}

The absence of infinitive attributes for self-experiments means that there
are no further restrictions to equilibrium fluctuations. Then the $G$-defect
entity can be connected with an internal temperature fluctuation $\delta T$.
The rational Gibbs thermodynamics has no such a $\delta T$, because $(%
\overline{\Delta T^{2}})^{1/2}=0$ is supposed, and new entities (besides the
particles) cannot be described there. von Laue [\cite{vonLaue,Landau} (\S
112)] suggested a thermodynamics for two independent pairs of variables
(entropy $S$ and temperature $T$, volume $V$ and pressure $p$) that allows
to calculate the average temperature fluctuation of a representative
subsystem:

\begin{equation}
\overline{\Delta T^{2}}=k_{B}T^{2}/C_{V},  \label{Eq.2.1}
\end{equation}%
where $C_{V}$ is the extensive heat capacity, $[C_{V}]=$ J/K, and defines
the size of the subsystem. The smallest representative (in the
time scale of the dynamic glass transition) contains one $G$ defect $-$
otherwise it would not be representative $-$, its characteristic volume is
therefore \cite{DonthJNCS82}

\begin{equation}
V_{mt}=k_{B}T^{2}\Delta _{mt}(1/c_{V})/\rho (\delta T)_{mt}^{2}
\label{Eq.2.2}
\end{equation}%
where the index $mt$ means the main transition, i.e. the dispersion zone of
the dynamic glass transition; the small letter $c_{V}$ is the specific heat
capacity, $[c_{V}]=$ J/kg$\cdot $K. The cubic root of $V_{mt}$ is called the
characteristic length, $\xi _{mt}=V_{mt}^{1/3}$. The characteristic volume
can be determined by dynamic heat capacity, e.g. \cite{Donth2001}, similar
results are obtained from dynamic compressibility, $\partial V/\partial
p(\omega )$. For high temperatures (and frequencies) a characteristic volume
of order the volume one molecule is obtained, for low temperature near the
conventional glass temperature $T_{g}$, the order of 100 molecules is
systematically reached \cite{Donth2003}. This behavior suggests that the
characteristic length is a reasonable quantity.

\subsection*{2.2 Gibbs prior consequences}

If a similar formula would be throwed together \cite{Donth1977} from Gibbs,
a thermodynamic compliance $\Delta c_{V}$ is obtained instead of a modulus ($%
\Delta (1/c_{V})$ in equation (\ref{Eq.2.2}),

\begin{equation}
V_{mt}^{\text{Gibbs}}=k_{B}T^{2}/\Delta _{mt}c_{V}\cdot \rho \cdot
\delta T^{\prime 2}  \label{Eq.2.3}
\end{equation}%
where $\delta T^{\prime }$ is now calculated from the $mt$ transformation
interval corrected by freezing-in effects. Using the same set of
experimental data as for equation (\ref{Eq.2.2}) \cite{Donth2003}, the $%
V_{mt}^{\text{Gibbs}}$ values are much larger than $V_{mt}$ from (\ref%
{Eq.2.2}) $-$ reflecting the Landau-Lifshitz subsystem size [\cite{Landau},
\S \S 1, 2, 35] based on statistical independence as defined by the Gibbs
distribution (or static correlation functions) $-$ and do not show a
systematic increase with falling temperature. This behavior suggests that
the construct (\ref{Eq.2.3}) is not a reasonable quantity.\newline

\subsection*{2.3 Experimentum Crucis for $G$ defect entities}

Today, the majority of thermodynamic colleagues does not believe in equation
(\ref{Eq.2.2}). To prove unambiguously the existence of such an entity, we
need another experiment that uses a definite, continuously varying length
testing the glass transition zone ($mt$) at different temperatures $T$ and
frequencies $\omega $, e.g. a wave vector $Q\approx2\pi $/length, and which length
can be compared with the characteristic length from thermodynamics. Static
Xray or neutron scattering is not suitable, because the appropriate contrast
in pure liquids is too low. \ More suitable is e.g. the dynamic neutron
scattering (DNS) in partly deuterated, otherwise pure liquids. But today, a
direct comparison with e.g. dynamic calorimetry (DC) is not possible because
there is a frequency gap \cite{Donth2001} between the former ($\gtrsim 10$
Mz) and the latter ($\lesssim 0.01$ Mz) method. According to an judgement of
leading people in DNS, e.g. Dieter Richter \cite{Richter2007}, and in DC,
e.g. Christoph Schick \cite{Schick}, the gap can be closed in some years, in 2012, say. If the lengths obtained from the two methods would be
consistent (and well outside the Gibbs formula), then this Experimentum
Crucis would be positive for the new entity.\newline
\newline

A $\log \omega -\log Q$ plot could be used for the comparison. The relevant
(diffusion, relaxation) part of the dynamic intermediate scattering function
gives a Levy extended-exponential time $\tau _{mt}(Q)$ \cite{Tyagi} that can
be transferred (via the Levy exponent $\alpha $ \cite{Donth2001}) in a
frequency $\omega (Q)$. From this, a raster \cite{Kahle} of DNS isotherms in
the above plot can be constructed. The ($\omega ,T$) points from DC can be
put in this raster, and the lengths from DNS-$Q$ values on the abscissa,
say, can directly compared with characteristic lengths from DC. It remains
to check the minor uncertainty, how density and entropy responses can be
compared.

\section*{3. Length relation between $G$ defects and $F$ speckles}

\subsection*{3.1 Shaping power of Levy distribution for $G$ defects}

Let us explain the concept of {\it shaping power of Levy distribution} in
more detail than above. Consider a common space in two dimensions: the $x$
axis be the natural space across the $G$ defect and the $\omega $ axis be
the frequency axis of the event space of probability. After a common scaling
(shorter modes are quicker, i.e. have a higher probability), the center $x=0$
is the middle of the defect with the larger frequency $\omega $ values (''$%
\omega \rightarrow \infty $'') and the periphery is the borderline region to
the neighbor defect with the lower $\omega $ values. Center and periphery of
the $G$ defect substitute thermodynamically a cooperatively rearranging region %
\cite{AdamGibbs}, an entity with low density contrast. In the
common space, the particular properties of the Levy distribution are
differently located. This is called shaping power. The preponderant
component \cite{Darling} with an influence of order of the tilt ($1-\alpha $%
) on the whole Levy sum of random variables, and the upper, fractal
hierarchy \cite{Bardou} of this ordered sum (by $n$) are located at the
center; while the damping factor $n^{-1/\alpha }$ of the ''lower hierarchy''
(large $n$) is arranged near the periphery in the entities where the sum breaks
down by diving below the neighbor defect. In other words, irrespective of
the low ''static'' density contrast, the borderline between the entities is
defined by the mutual disturbation of the differently centered (in space)
shaping powers of neighboring $G$ defects; the borderlines define the
characteristic volume (\ref{Eq.2.2}). Since the Gauss distribution ($\alpha
=2$) has no particular properties such as hierarchy etc., a shaping power
cannot be defined for Gauss and Gibbs: The shaping power is for the dynamic
heterogeneity.\newline
\newline

The Levy distribution density $p(\omega )$ is thus realized by a spectral
density $x^{2}(\omega )$ in the measure $d\omega $ with $dx=x^{2}(\omega
)d\omega $. \ For large frequency at the $G$ defect center we find
fractality,

\begin{equation}
p(\omega )\sim \omega ^{-1-\alpha }. \label{Eq.3.1}
\end{equation}%
The center is then connected with the local break through of the mobility
(Levy instability). The preponderant component corresponds to the diffusion
step of the molecule through a cage door of neighbor molecules in the
center. As known from probability theory \cite{Darling}, for $\alpha <1$ the
Levy exponent $\alpha $ is often the only important parameter of the limit
distribution. Theoretically, the Levy exponent $\alpha $ must be realized
during a renormalizing limit process of the sum, $n\rightarrow \infty $. The
Levy exponent $\alpha $ depends on a plurality index $q$ that depends for a
given defect on the kind of disturbance / response, on temperature, and
pressure. The reason for plurality is that the numbers $n$ for the Levy sum
of a $G$ defect are not so large as e.g. the huge number of filter elements
for the renormalization of charge and mass of elementary particles, where a
pure huge-number approximation seems to be a reasonable approach (cf. \cite%
{this part}, sections 5.2 and 5.3 there).\newline
\newline

As a consequence of this picture, the local diffusion is determined by the
fractality, as observed by DNS \cite{Arbe}. A sublinear {\it Levy diffusion}
is obtained as

\begin{equation}
\omega \rightarrow \omega ^{\alpha }\text{ , }\tau \sim \xi
^{2/\alpha } \label{Eq.3.2}
\end{equation}%
having the Cauchy limit $\alpha =1$ for {\it normal diffusion}, $\tau \sim
\xi ^{2}$ ($\xi $ being the relevant length).\newline
\newline

Remark. Very general reasons are necessary to apply the above shaping-power
scenario to a cosmological initial liquid. Example. The representativeness
theorem \cite{Donth2001} shows that the Levy dynamic heterogeneity of the
dynamic glass transition ($mt$) can even be derived from arguments of
classical thermodynamics alone. This theorem says that the dynamic
compliances have Levy-distributed times with a Levy exponent $\alpha \leq 1$%
, whereas the corresponding ''static'' spatial correlations have a Gauss
distribution ($\alpha =2$). Roughly: The Levy distribution follows from a general limit theorem of probability theory (excluding others than Gauss and
Levy, cf. Feller in \cite{Darling}), since fluctuation (variance) and expectation is infinite due to the
Levy instability in the cage. This is a new entity, thermodynamically a minimal representative with one $G$ defect
(representative for the whole system, cf. the equation (\ref{Eq.2.2})
arguments); the infinities follow from the negative pressure inside,
especially $\alpha \leq 1$ follows from the infinitive expectation of the
diffusion step. Since the probability density of classical systems is
symmetric, the characteristic function is proportional to $\exp (-a\left|
t\right| ^{\alpha })$ leading directly to the experimental Kohlrausch
function. The use of compliance (instead of a modulus) follows from the
relation to the additivity of the ordered Levy sum via the FDT as measuring
equation.

\subsection*{3.2 Largest causal region ($F$ speckle) from Levy diffusion ($G$
defects)}

One of the greatest surprise in classical glass transition research was the
discovery that the Debye B\"{u}che inhomogeneities in frozen glasses \cite%
{Debye} correspond to a slow dispersion zone. These ''Fischer modes'' $\phi $
have relaxation times e.g. 10$^{7}$ times longer and mode lengths e.g. 50
times larger than those of the dynamic glass transition ($mt$) \cite%
{Fischer,Patkowski}. The corresponding Fischer speckles contain, according
to the length ratio $\phi /mt$, a large number $N_{G}$ of $G$ defects. The
origin of such large numbers is the next task.\newline
\newline

We assume $-$ against wild structural speculations in the
literature, e.g. Fischer clusters $-$ that the Fischer modes are a
direct consequence of the dynamic glass transition $mt$ without
any structural input \cite{Donth2001}. Then we have to look for a
physical $G$ defect process that eats its way through large space
$\xi $ and time $\tau $ ranges: a Levy diffusion process equation
(\ref{Eq.3.2}). We expect, therefore, that the tilt ($1-\alpha $)
with the Levy exponent $\alpha $ defines the large number $N_{G}$
(figure 1).

\begin{figure}[here]
\begin{center}
\includegraphics[width=0.3\linewidth]{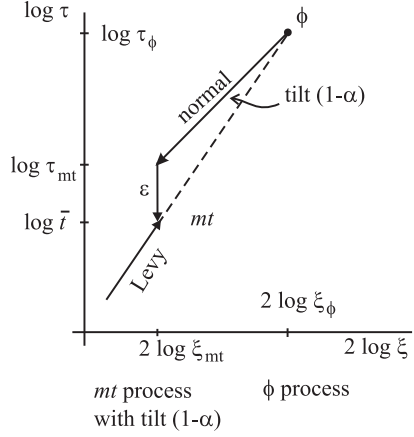}
\caption{Tilt construction for Fischer $\phi $ process length
($\xi
_{\phi }$) and time ($\tau _{\phi }$) in a diffusion plot, log$\tau $ vs $%
\log \xi ^{2}$, using the relevant tilt ($1-\alpha $) for the
Glarum Levy
defects ($G$ defects) of the main transition ($mt$). Forward = Levy ($%
\nearrow $), backward = normal ($\swarrow $), cf. figure 2 below.
\label{fig1}}
\end{center}
\end{figure}

Forward, how probes the fast $G$ defect process the slow $\phi $
process?\ The $G$ defects have a large density, in classical
liquids ($1/V_{mt}$) from equation (\ref{Eq.2.2}) of order
$1/$nm$^{3}$. The molecular diffusion is therefore step by step
with short times $\tau _{mt}$ and short lengths $\xi _{mt}$, from
one $G$ defect to the next neighboring one along a diffusion line
(figure 2a). We find therefore the Levy diffusion (\ref{Eq.3.2})
as observed \cite{Arbe}, along to full distances of the $F$
speckles by ''analytic continuation'' with the sublinear slope.

\begin{figure}[here]
\begin{center}
\includegraphics[width=0.25\linewidth]{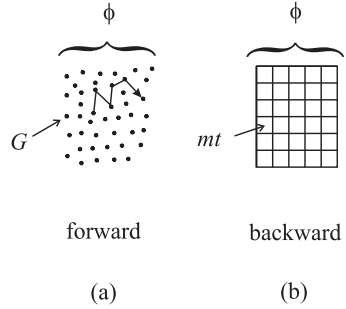}
\caption{Comparison of forward ($G\rightarrow \phi $) and backward ($%
\phi \rightarrow mt$) probes. Details see text. \label{fig2}}
\end{center}
\end{figure}

Backward, how probes the slow $\phi $ process the fast $G$ defect process?
In the large $\phi $ time order $\tau _{\phi }$, the fast dynamics of the
small $mt$ subsystems are mutually statistically independent (for $\tau
_{mt}\ll \tau _{\phi }$). The many $mt$ subsystems allow a quasi
thermodynamic treatment. The thermodynamic variables $X$ have a finite
variance. The relaxation can then be described by $\Delta \dot{X}=-$ const $%
\Delta X$, with a positive constant. This gives the exponential decay
leading via a Cauchy law ($\alpha =1$) to normal diffusion in equation (\ref%
{Eq.3.2}).\newline
\newline

Consequently, we find the space and time scales for the $F$ speckles ($\xi
_{\phi },\tau _{\phi }$) at the upper ($\phi $) corner of the triangle in
figure 1: The Levy diffusion, to be effective, must be faster than the
normal diffusion. In larger scales, beyond the $\phi $ corner, the
construction is exhausted, which means that the $F$ speckle defines a
largest causal region in the classical liquid.\newline
\newline

To get a formula, we start from the triangle basis $\varepsilon $. This is
the logarithmus-of-time difference between the normal and Levy diffusion at
the $G$ defect length scale $\xi _{mt}$. The Levy diffusion is at high
''fractal'' frequencies near the $G$ defect center at the cage door, i.e. at
some $\bar{t}$ in the short-time tail of the $mt$ process. The normal
diffusion, however, is near an average of time (in logarithmic measure), $%
t_{av}$:

\begin{equation}
\varepsilon =\log _{10}(t_{av}/\bar{t})_{mt} \label{Eq.3.3}
\end{equation}%
i.e. $\varepsilon $ is of the order $1/\alpha $, the width of the response
function on the $\log _{10}\omega $ axis (figure 3.7b of \cite{Donth2001}).%
\newline
\newline

Figure 1 shows that the length and time ratios ($\phi /mt$) are mainly
determined by the Levy exponent tilt of the $G$ defect dynamics, ($1-\alpha $%
). Using $\tau _{mt}=t_{av}$, we get the slope difference (forward minus
backward) as ($1-\alpha )/\alpha $ and the ratios as

\begin{equation}
\log _{10}\frac{\xi _{\phi }}{\xi _{mt}}=\frac{\alpha \,\varepsilon }{%
2(1-\alpha )}\text{ and }\log _{10}\frac{\tau _{\phi }}{\tau _{mt}}=\frac{%
\alpha \,\varepsilon }{1-\alpha }. \label{Eq.3.4}
\end{equation}%
Both ratios tend to infinity for vanishing tilt, $1-\alpha \rightarrow 0$
(or $\alpha \rightarrow 1$, Cauchy). Large ratios can therefore easily be
obtained for small tilts, also such ones as observed in the classical
liquids \cite{Donth2001}. Example. The number $N_{G}$ of $G$ defects as
claimed by the Fischer speckles as largest causal region in $d$ dimensions is

\begin{equation}
N_{G}=%
%TCIMACRO{\QOVERD( ) {\xi _{\phi }}{\xi _{mt}}}%
%BeginExpansion
{\xi _{\phi } \overwithdelims() \xi _{mt}}%
%EndExpansion
^{d}=10^{\frac{d\alpha \varepsilon }{2(1-\alpha
)}}(=10^{\frac{3\alpha \varepsilon }{2(\alpha -1)}}\text{ for
}d=3). \label{Eq.3.5}
\end{equation}%
For \{$d=3$, $\varepsilon =1$, $1-\alpha =0.13$\} we get $N_{G}=10^{10}$;
this $N_{G}=10^{10}$ value can also be obtained e.g. for $d\cdot \varepsilon
=0.83$ and $(1-\alpha )=0.04$, and so on.\newline
\newline

For cosmological applications (section 6), we should discuss the relation of
the $G$ defect tilt ($1-\alpha $) to the scalar index tilt ($1-n_{S}$) from
the power law fit of the density fluctuation spectrum. It seems useful,
however, first to devote the next sections 4 and 5 to gravitation and
inflation of our model.

\section*{4. Graviton as torus for a charge, randomly constructed from four $%
\protect\vartheta $ segments of the hidden charge model}

Our model \ does not have an eigensolution for gravitons. If the graviton is
some type of charge (gravitation as a charge is discussed by Feynman \cite%
{Feynman}), and if a charge is characterized by at least one
$S^{1}$ torus in the model, then an extra torus beyond the
eigensolutions must be constructed: A circle $S^{1}$ from four
$\pi /2$ segments of $\vartheta $ may be linked by four random
connections called spots ($\bullet $ in figure 3). The mass
concentration into galaxies can give spots in the particle
environment, if the culminating-point filter construction from
large to small filter elements is considered as some map of the
universe in the Mach's principle layer of the tangent objects
(\cite{this part}, table 2 of appendix C there). E.g. the central
black holes of galaxies may give the randomly selected linking
spots in the ''dollhouse universe'' around any particle (section
7).

\begin{figure}[here]
\begin{center}
\includegraphics[width=0.25\linewidth]{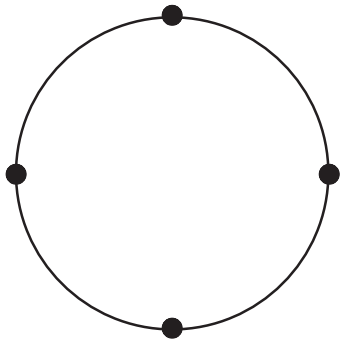}
\caption{Graviton torus from four $\pi /2$ \ \ $\vartheta $
segments in the tangent objects of any particle. \label{fig3}}
\end{center}
\end{figure}

The number of spots is equal to the number of galaxies in the universe, $%
N_{G}\approx 10^{10}$. Assuming some statistical independence in the random
selection of the four spots, we get a probability for the graviton of about

\begin{equation}
P\text{ (graviton) }\approx 1/N_{G}^{4}. \label{Eq.4.1}
\end{equation}%
This probability may be compared with the probability of order $P=1$ to find
a charge torus for the other particles participating in an interaction
considered. This means that the gravitation is, for given ''distances'',
weaker by a factor of order 10$^{40}$ than the electroweak and strong
interaction.\newline
\newline

A review of our model constructions for particles (eigensolutions [\cite%
{this part}, section 3], for common eigensolutions [\cite{this part},
section 5], for exchange boson construction (\cite{this part}, appendix C),
and for the dollhouse universe (section 7, below) seems to show that now the
application of torus constructions for particles is exhausted (table 1). Our
model is now complete.

%TCIMACRO{\TeXButton{B}{\begin{table}[tbp] \centering}}%
%BeginExpansion
\begin{table}[tbp] \centering%
%EndExpansion
\caption{List of $S^1$ torus topologies defining charges of our
model.\label{Tab.1}}%
\begin{tabular}{ccc}
&  &  \\
torus & name & application \\
coordinates &  &  \\ \hline
&  &  \\
($\varphi _{1},\varphi _{2}$) & $S^{1}$ in the $S^{3}$ & ($\oplus $, $%
\ominus $) \\
& Heegaard tori & electrical \\
& (\cite{this part}, figure 6 there) & charges \\
&  &  \\
$\tau $ & $S^{1}$ tori of & photon \\
& the hidden & identity \\
& space $S^{1}\times S^{3}$ &  \\
&  &  \\
$\vartheta $ & dollhouse & graviton \\
from four & tori, & identity \\
$\pi /2$ segments & figure 3 &
\end{tabular}%
%TCIMACRO{\TeXButton{E}{\end{table}}}%
%BeginExpansion
\end{table}%
%EndExpansion
Via an adapted van der Waerden (vdW) construction \cite{Donth1988}, the
gravitation torus of figure 3 defines a metric tensor $g_{ik}(x)$ with spin $%
S=2$ on the reality space from $M_{4}$ tangent objects. Because the
gravitation is so weak and stems from the far universe, we can speak about a
{\it shadow metric over the reality }$M_{4}${\it \ space}, if the particles
or mass concentrations are widely isolated in a vacuum (''much vacuum'').

\section*{5. Hadronic inflation of cosmological expansion}

In our model, quantum-mechanically behaving cold hadrons are not realized
before the expansion of the universe reaches the length scale of order $%
R\approx 1$ fermi. Moreover, in this scale the first black holes are formed
in the center of $G$ defects.\newline
\newline

The empirical mass scale $m_{0}$ above the neutrinos is from MeV to TeV,
without the electrons we find a scale from GeV to TeV; inclusive the stable
atoms. In sufficiently large $M_{4}$ tangent objects, we have a Compton
wave-length of

\begin{equation}
\lambda _{c}=2\pi \hbar /m_{0}c\approx \frac{1.24\text{ fermi}}{m_{0}/\text{%
GeV}}\text{ , 1 fermi }=10^{-15}\text{m.} \label{Eq.5.1}
\end{equation}%
Assuming that only the heavy particles are of cosmological influence for a
cold expansion, then the energy scales for quantum flavor dynamics (QFD) and
chromo dynamics (QCD) have length scales of

\begin{eqnarray}
\Lambda _{\text{QFD}} &\approx &250\text{ GeV , }\lambda _{\text{QFD}%
}\approx 0.005\text{ fermi} \label{Eq.5.2} \\
\Lambda _{\text{QCD}} &\approx &150\text{ MeV , }\lambda _{\text{QCD}%
}\approx 8.3\text{ fermi .}  \nonumber
\end{eqnarray}%
Shortly, we call this generous region ''hadronic'':

\begin{equation}
100\text{ MeV ... 250 GeV , 0.005 fermi ... 10 fermi}
\label{Eq.5.3}
\end{equation}%
with $\lambda _{0}=1$ fermi $\approx 1$ GeV and the proton mass $m_{p}$ as
typical parameters. For high temperatures, $k_{B}T\gg m_{0}c^{2}$, the de
Broglie wave lengths are much shorter. This is called ''hot''; our hadronic
scale (\ref{Eq.5.3}) is then called ''cold''. We try to describe the
expansion of the universe as a cold phenomenon, because our model mass scale
is confined from above by existential instability (generalisation of figure
5 in \cite{this part}).

\subsection*{5.1 Cold inflation}

Consider in advance a box with a length $l$ ($l$ in meters). Without $x/\psi
$ or $M/E$ separation, the box {\it must} be filled with particles that {\it %
must} be hot for small boxes $l<\lambda _{0}$ (because of the quantum
mechanical uncertainty relation). With $x/\psi $ or $M/E$ separation,
however, an empty small box is possible: A culminating-point particle at $x$
supposes the existence of small filter elements $l^{\prime }<l$. This is not
possible for cold particles in case of $l<\lambda _{0}$, because then
smaller elements ($l^{\prime }<\lambda _{0}$) are not possible due to the
upper limit of the hadronic region (equation (\ref{Eq.5.3}), see also figure
6 of \cite{this part}). The uncertainty relation follows from $\psi (x)$
assuming the possibility of existence of a particle at $x$ as the basic
prerequisite. This {\it must} be assumed for a manifold diffeomorphism
(without $x/\psi $ separation), but this is excluded for cold particles in
the too small boxes ($l<\lambda _{0}$) from our independent
culminating-point particles (with $x/\psi $ separation).\newline
\newline

Let us now specify the concept of {\it filter elements} sometimes used
above. Since a filter is considered as some kind of a map from the hidden $H$
in the reality $M$ space, it contains things from the origin $H$ (e.g.
vacuum elements) and the image $M$ (metric for $M_{4}$ or $E^{4}$ tangent
objects). Our mathematical filter sets contain, therefore, a certain size
variation and get some relation to a Cauchy filter (figure 1b of \cite{this
part}). The filter sets of vacuum elements resulting from existential
instability are therefore called filter elements.\newline
\newline

The method used for our inflation scenario is the {\it cutoff filter} of
figure 4. The filter elements on the $M_{4}$ tangent objects for cold model
hadrons (in the vdW spinor space) should not become smaller than the
hadronic scale $\lambda _{0}$, because then the Compton wave length (as
condition for their cold quantum mechanical existence) is not available for
virtual filter elements in the too small universe ($R\ll \lambda _{0}$).
Such cold quantum mechanical particles cannot exist as points there, but
only as larger constructs of the model, hidden by the eigensolutions on the $%
H$ space ($S^{1}\times S^{3}$), and lurking there for their
convergence to particles in the reality space for $R\gtrapprox
\lambda _{0}$. In other words, for $R\ll \lambda _{0}$ there is no
room enough to realize a cold quantum mechanical particle (a
hadron) in the reality space (cf. with the box discussion above).

\begin{figure}[here]
\begin{center}
\includegraphics[width=0.4\linewidth]{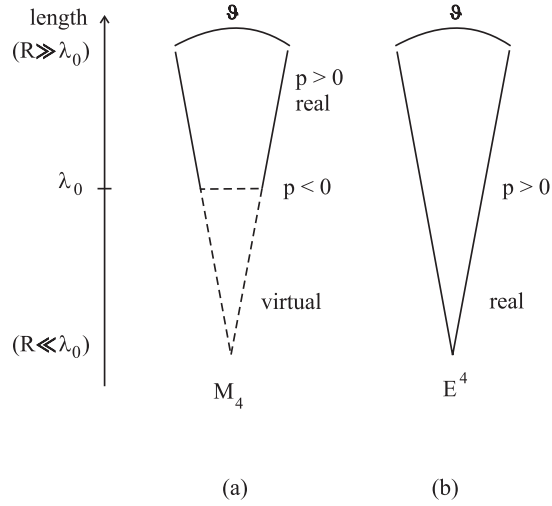}
\caption{Filter construction on $M_{4}$ and $E^{4}$ tangent
objects. {\bf a} (left). For cold particles we find a cutoff
filter for $R>\lambda _{0}$. The cutoff at $R\approx \lambda _{0}$
has a negative pressure from creation of many hadrons during
expansion. {\bf b} (right). For non-isolated dark matter particles
in an $E^{4}$ Euclidean environment, the filter is not cut and we
find always positive pressure, $p>0$. If, however, because of
expansion, dark matter particles become isolated, $-$ i.e. if they
will be imbedded in Minkowski space $-$ then they are also to be
treated by the left hand side cutoff scenario (a). \label{fig4}}
\end{center}
\end{figure}

This holds also for 2-class particles (e.g. dark matter or prebaryons) with
''internal'' $E^{4}$ tangent objects, if they are isolated by imbedding in
an ''external'', sufficiently large Minkowski space from $M_{4}$ tangent
objects of the other particles or other participants of common culminating
points. This does not hold, however, if they are not isolated, i.e. if they
are living in an $E^{4}$ environment from tightly neighboring particles of
2-class eigensolutions, e.g. from dark matter particles. The missing time in
$E^{4}$ does not allow to define the Planck constant $\hbar $ needing the
seconds. Then the filter is not cut (figure 4), and cold dark matter
particles (and prebaryons) can also exist at high density for $R\ll \lambda
_{0}$. Dark matter particles are preferred in the small universe.\newline
\newline

In other words, if the cosmological expansion of a cold initial dark-matter
universe crosses the hadronic length scale, $R\approx \lambda _{0}$, then
many hadrons (and hadron pairs) are formed which were absent before in the
reality-space of the universe. This is a typical situation for negative
pressure, $p<0$ (figure 4 again). [Remember: Thermodynamically, the pressure
$p$ is one of the variables that are necessary for maintaining the
considered equilibrium state. For representative subsystems with length $l$,
here for $l\ll R$, this pressure $p$ would be the ''external'' pressure.
There is a widespread misunderstanding, that negative pressure can be
defined by radially escaping particle trajectories. Such behavior would also
be typical for positive external pressures $p^{\prime }$ with $0<p^{\prime
}<p$.] During a period of forming many new particles, controlled by the
Lagrangian, the (stationary) state can only be maintained, if the pressure $%
p $ is negative, $p<0$, and the system (our universe), therefore, is
expanding.\newline
\newline

This expansion stage directly driven by hadronic forces is called {\it %
hadronic inflation}. If this is connected with a ''hadronic cosmological
constant'', suitably compared to gravitation, we would get the factor 10$%
^{40}$ of equation (\ref{Eq.4.1}). The cold hadronic inflation is therefore
an extremely violent stage of expansion without hot X particles from a gauge
unification at 10$^{16}$ GeV.

\subsection*{5.2 Stages of the hadronic inflation}

The resulting scenario for the hadronic inflation is characterized
by the following concepts (figure 5). The infinitely large liquid
system (''primeval soup'') is called ''original'', because it
exists also before the universe comes into being as consequence of
a Linde type of fluctuation. The developing small ($R\ll \lambda
_{0}$) universe is called ''initial'', because it is related to
its beginning and its independence; it is a first step in a series
of processes, developing from a Linde fluctuation in a maximal
causal region claimed by an $F$ defect of the original liquid. The
violent increase near $R\approx \lambda _{0}$ is called ''hadronic
inflation'' as defined above. Immediately after this violent
hadronic pulse, the universe cannot be controlled by the too weak
gravitation. Later, however, for $R\gg \lambda _{0}$, the
expansion control is captured by gravitation and is then called
''primordial''.

\begin{figure}[here]
\begin{center}
\includegraphics[width=0.4\linewidth]{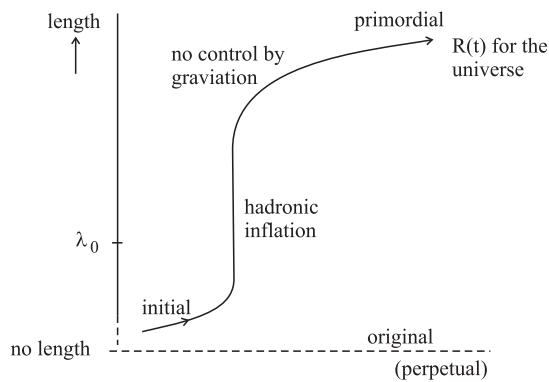}
\caption{Terminology of the hadronic inflation scenario.
\label{fig5}}
\end{center}
\end{figure}

The stages of the scenario are discussed according to our model with respect
to the crumbled space (mix of $M_{4}+E^{4}$), to gravitation, and to
expansion.\newline
\newline

{\it Original state}. Assuming a large volume of the original liquid, we get
point particles from 1-, 2-, and 3-class eigensolutions via the
non-virtuality of the small filter elements, e.g. photons, hadrons, and dark
matter point particles. The high density corresponds to a crumbled space of $%
M_{4}$ and $E^{4}$ tangent objects. Although $G$ defects seem to bee
possible, we get no gravitation, because their number $N_{G}$ tends to
infinity in the large volume ($1/N_{G}\rightarrow 0$) and, additionally, the
links for gravitons (${\Large \bullet }$ in figure 3) are expected to be too
weak because of their weak contrast. Dynamic $F$ speckles as a consequence
of $G$ defects should still define maximal causal regions for Linde
fluctuations.\newline
\newline

{\it Initial liquid}. Since only the dark matter particles and prebaryons
remain points for $R\ll \lambda _{0}$, figure 4b), their fraction in the
expanding Linde fluctuation should be large. We expect a fluctuation
(Boltzmann zacke) with many dark matter particles in the high-density
liquid, and with no isolated hadrons because of virtuality in their cut off
filter. If there are cold photons (no $\lambda _{c}$) but too weak for an
imbedding $M_{4}$ space, we extremely get an Euclidic $E^{4}$ space, or at
least a crumbled space with a high $E^{4}$ fraction. In the extremum, time
seems to be standing still, and we find only a ''shape of time'', some
succession of fourdimensional Euclidic $E^{4}$ spaces. The reason for the
expansion in this stage is therefore highly speculative also in the frame of
our model and rests on a certain cold photon fraction. E.g., one possibility
would be the vanishing of the original $M_{4}$ particles (hadrons) due to
their transfer into the virtuality in such a Boltzmann zacke (figure 4a).
This could mean a positive pressure $p^{\prime }$, $0<p^{\prime }<p$, in the
universe for $R\ll R_{0}$.\newline
\newline

The total topology of the whole universe from the $F$ speckle may be formed
in this expansion stage. $R$ is the radius of the emerging universe in
meters. The topology may be induced from the two hidden Heegaard tori of the
charge symmetry and from imbedding of $R$ in a crumbled space with
dominating fourdimensional $E^{4}$ tangent elements. This is, I think, a
situation that favors a total $S^{3}$ sphere instead of a $D^{3}$ disk,
irrespective of local curvature fluctuations inside.\newline
\newline

{\it Hadronic inflation}. For $R\approx \lambda _{0}$ the hadrons come into
the play. This generates the negative pressure leading to a violent
expansion transferring the dark matter particles in the isolation. They are
then imbedded in the Minkowski space allowing their quantum mechanical
behavior for $R\gg \lambda _{0}$. \ If the $G$ defects from the zacke are
intensified to become the ${\Large \bullet }$ links \ for the gravitons,
then gravitation is switched on during the inflation. This intensification
could be possible by their transfer to black holes (section 5.3). [This
would mean that the galaxies come finally from the shaping power of Levy
distributions, section 3.1.]\newline
\newline

{\it Primordial stage}. If we put on a material variant
then the emerging particles drive the expansion, not a vacuum field: this
stage develops from the motion of interacting ''hadronic'' particles. In the
inflation stage ($R\approx \lambda _{0}$) and for high particle density in
the cold universe, the gravitation has no or at most a minor influence (10$%
^{-40}$, equation \ref{Eq.4.1}). In the further expansion, we get increasing
distances between the particles and particle congregations. For electric
neutrality, the vacuum gets influence via the gravitation (section 5.3). A
control by gravitation, however, is only possible for not too small $R$. The
expansion can then be captured by the Friedmann Lamaitre (FL) equations,
containing besides the Newtonian gravitational constant $G_{N}$ the
mysterious cosmological parameter for gravitation, $\Lambda $ (section 7.2).%
\newline
\newline

This material variant has no superluminar particle velocities and is based
on special relativity in the Minkowski space from the $M_{4}$ tangent
objects. We have no serious horizon exit and reentry problems during and
after the violent pulse from cold inflation. Our material variant is an
alternative to the conventional vacuum variant: The superluminal velocities
there may result from requirements to the FL equations, e.g. stationary
during inflation: $1/H=$ constant, density $\rho _{V}=$ const, and a field $%
\phi $ for the vacuum $V$. Let us repeat: It is the field $\phi $ that is
considered in the vacuum variant, not cold particles with mass in the
material variant as used here.

\subsection*{5.3 Cold Hawking unification?}

The Hawking temperature \cite{Hawking1975} for black holes,

\begin{equation}
k_{B}T_{H}=\hbar c^{3}/8\pi \,G_{N}\,M, \label{Eq.5.4}
\end{equation}%
connects gravitation ($G_{N}$), hole mass from particles ($M$), photon ($c$)
and thermodynamics ($k_{B}$). Equation (\ref{Eq.5.4}) is sometimes
considered as the basis for an unification of gravitation with the other
interactions. Using, however, conventional gauge invariance principles, the
unification with gravitation is usually obtained in a hot universe that is
characterized by temperatures or energies of order the Planck mass ($m_{%
\text{Pl}}=(\hbar c(G_{N})^{1/2}\approx 10^{19}$GeV). We try to find a cold
alternative in the frame of our model.\newline
\newline

Consider the $G$ defect as representative thermodynamic entity (from which
the one $k_{B}$ for the Hawking temperature comes). The physical process for
the equilibration of temperature is now the exchange of particles (mass $%
m_{0}$) at the horizon of the black hole. $-$ We consider first the particle
concentration inside the $G$ defect. In ''cosmological approximation'', the
extensive entropy is equal to the number of particles in the mass
concentration of the $G$ defect, identified with $M$:

\begin{equation}
S_{m}/k_{B}\approx N_{\text{particles}}\approx M/m_{0}\propto M.
\label{Eq.5.5}
\end{equation}%
The corresponding Bekenstein Hawking entropy \cite{Bekenstein} for the black
hole is

\begin{equation}
S_{A}/k_{B}=4\pi \,G_{N}\,M^{2}/\hbar c\propto G_{N}\,M^{2}\propto
M^{2}. \label{Eq.5.6}
\end{equation}%
We ask where the two formulas cross each other ($T^{X},\lambda
^{X}=r_{S}^{X},M^{X}$), equation (\ref{Eq.5.5}) for the $G$ defect
mass concentration without a black hole, and (\ref{Eq.5.6}) for a
black hole formed by this mass concentration of the $G$ defect
(figure 6).

\begin{figure}[here]
\begin{center}
\includegraphics[width=0.3\linewidth]{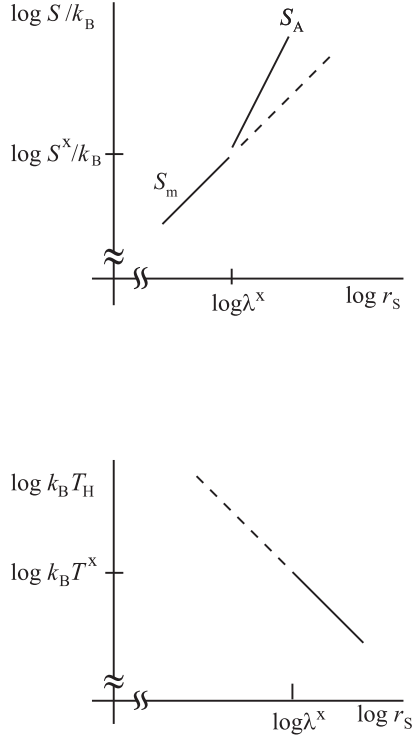}
\caption{Crossing of particle entropy of a $G$-defect mass
concentration ($S_{m}$) with a horizon area entropy ($S_{A}$) for
a corresponding black hole. The drawn lines are from an entropy
maximum principle, the broken lines are extrapolations. $\lambda
^{X}$ is the cross length scale related to the Schwarzschild
radius of the black hole, $r_{S}$.
Below $\lambda ^{X}$ of order the hadron Compton length (equation (\ref%
{Eq.5.9}) below), the particle concentration of the $G$-defect
with no black hole is the optimum. \label{fig6}}
\end{center}
\end{figure}

Using the Schwarzschild radius

\begin{equation}
r_{S}=2G_{N}\,M/c^{2} \label{Eq.5.7}
\end{equation}%
we get from (\ref{Eq.5.5}) and (\ref{Eq.5.6}) for the crossing of figure 6

\begin{equation}
M^{X}=\lambda _{0}c^{2}/G_{N}8\pi ^{2}\text{ , }\lambda
_{0}=h/m_{0}c, \label{Eq.5.8}
\end{equation}%
where $m_{0}$ is the mass of particles considered. Then we have

\begin{equation}
r_{S}^{X}=\lambda _{0}/4\pi ^{2}\text{ and }k_{B}T^{X}=\pi \hbar
c/\lambda _{0}. \label{Eq.5.9}
\end{equation}%
Schwarzschild radius and Hawking temperature at the formation of first black
holes are in the hadronic region. The number of particles with mass $m_{0}$
is

\begin{equation}
M^{X}/m_{0}=hc/8\pi ^{2}m_{0}^{2}G_{N}\propto
1/m_{0}^{2}G_{N}\text{.} \label{Eq.5.10}
\end{equation}%
\newline
For protons ($m_{0}=m_{p}$) we get a mass $M^{X}$ of the first black holes
of order 10$^{10}$kg, for dark matter particles ($m_{0}\approx 100m_{p}$) of
order 10$^{8}$kg. This mass contains about 10$^{38}$ protons or 10$^{34}$
dark matter particles: gigantic numbers.\newline
\newline

The reason for such numbers is the small (10$^{-40}$) gravitational constant
in the denominator of equation (\ref{Eq.5.10}). A black hole can only come
if the particle numbers (and their density) are gigantic. We see also that
any galaxy gets a central black hole during inflation.\newline
\newline

Remark. There is no danger in the LHC experiment that such cosmological
black holes can artificially be generated. The corresponding $E=mc^{2}$
energy is of order 10$^{46}$eV (calculated with protons), whereas LHC gets
maximally 10$^{15}$eV. Higher dimensions as from string theory for small
black holes are excluded in our model. Our earth experiments cannot generate
a second central black hole in our galaxy.\newline
\newline

The crossover mass (\ref{Eq.5.8}) for protons (10$^{10}$kg) is of order 10$%
^{-20}m_{\odot }$ ($m_{\odot }$ is the sun mass) and is much smaller than
the mass of the today central black holes of the galaxies (10$^{10}m_{\odot
} $). There are two reasons for this smallness in our scenario. 1. The first
black holes from mass concentration are in the center of $G$ defects smaller
than the $G$ defects themselves. In a liquid, the center is at the fractal
tails of their Levy distributions at large frequencies $\omega $; the
preponderant component of the Levy distribution is a part of the mass
concentration. In the $E^{4}$ space, this corresponds to large wave vectors $%
k$ (section 6). This means that the first black holes are formed at a small
part not only of $G$ defects, but also a very small part of the much larger $%
F$ speckles defining $R$ of the universe. The first black holes are formed
at diameters $R$ larger than $\lambda _{0}$, i.e. at a later stage of
hadronic inflation. 2. Since $T_{H}$ is decreased during the inflation
(figure 6), the growing black hole enters at some time a nonequilibrium
state, and the further growth becomes irreversible. This is a
self-amplifying process, because larger black holes become stronger spots $%
{\Large \bullet }$ in figure 3 which strengthens to linking of $\vartheta $
graviton segments increasing the gravitational constant, $G\rightarrow G_{N}$%
. There is much enough material left for forming of the galaxy structure. A
structure can also be formed in the early primordial state where essential
ingredients such as mass concentration and central black holes are already
available.\newline
\newline

In formulas, the results can bee summarized as

\begin{eqnarray}
r_{S} &\ll &\lambda _{0}:S_{m}\gg S_{A}\text{ , }T_{H}\gg \lambda _{0}\text{
, equilibrium,} \label{Eq.5.11} \\
r_{S} &\gg &\lambda _{0}:S_{m}\ll S_{A}\text{ , }T_{H}\ll \lambda _{0}\text{
, non-equilibrium.}  \nonumber
\end{eqnarray}%
If the optimum is really defined by the maximum of entropy from the
Boltzmann probability, $S=k_{B}\ln W_{B}$, we conclude

\begin{enumerate}
\item[a)] The first black holes of the expanding universe are formed in the
spatial center of the $G$ defects during later stages of the hadronic
inflation, (where $r_{S}\approx \lambda _{0}$ is reached).

\item[b)] These black holes are formed in thermodynamic equilibrium; later
in the larger expanding universe, the black holes fall out of the
equilibrium, and their growth becomes irreversible and is mechanically
controlled by the Newtonian gravitational constant $G_{N}$.
\end{enumerate}

In our model \ scenario, gravitation is unified with the other three
interactions insofar as they are relevant to our cold hadronic inflation:
via the Hawking temperature equilibration in the hadronic scale $\lambda
_{0} $ of the expansion, $R\approx \lambda _{0}$. This is called ''cold
Hawking unification''. The gravitational interaction comes into being not
before the inflation: Gravitation is first realized by strengthening of
torus links ${\large \bullet }$ for the graviton torus of figure 3.\newline
\newline

From a methodical point of view, the cold unification corresponds to some
calibration of mass values for leptons and hadrons. The filter construction
( \cite{this part}, section 5.3 there) calculates sharp mass ratios, e.g. $%
\log (m_{0}(2)/m_{0}(1))$, that can be related e.g. to $m_{0}(1)$ as
electron or to the proton mass. ''Absolute'' values can be obtained by
comparison with gravitation in the primordial state of the universe. The
gravitational constant $G_{N}$, however, has some fundamental randomness
from the number $N_{G}$ of $G$ defects inside the Fischer speckle (equation (%
\ref{Eq.4.1}) and figure 2b leads to the relative order of $1/\sqrt{N_{G}}%
=10^{-5}$ for the fluctuation variance), whereas e.g. the parameters of the
electroweak interaction are sharp from the huge numbers of the filter
construction.

\section*{6. Relation between tilt ($1-n_{S}$), number of galaxies in the
universe ($N_{G}$), and CMB temperature fluctuation amplitude ($\Delta T$)}

The way from a Levy tilt ($1-\alpha $) in the original liquid to the tilt ($%
1-n_{S}$) of cosmic density fluctuation in the primordial stage is discussed
in this section. First three model priors:\newline

{\bf a.} In the classical liquid, the number of $G$ defects ($N_{G}$) in the
maximal causal region ($F$ speckle) can bee estimated using the relevant
Levy exponent tilt ($1-\alpha $) with $\alpha =1$ for the classical Cauchy
diffusion (equation (\ref{Eq.3.5}), figure 1). Small tilts ($1-\alpha $)
correspond to large numbers $N$. The estimation is based on an analytic
continuation of the Levy diffusion from the $G$ scale to the $F$ scale
(figure 2).\newline

{\bf b.} The initial liquid before the hadronic inflation is dominated by
Euclidic $E^{4}$ tangent objects (section 5.1). To get a tilt in the initial liquid, the
shaping power from a classical diffusion Levy instability by free volume may
be substituted by the shaping power from an $E^{4}$-potential Levy
instability in the initial liquid. The elliptic type of interaction between
dark matter particles favors high densities in $E^{4}$ instead high
frequencies from large local free volume in classical liquids.

{\bf c.} The first black holes emerge from a mass concentration in the
center of $G$ defects of the initial liquid (section 5.2).

Now three changes of the tilt. First change. During the way from the
original classical liquid in $M_{4}$ to the initial liquid in $E^{4}$ the
time is lost (Hawking's shape of time). Instead of a frequency $\omega $ for
diffusion (besides three spatial dimensions) we have then a fourth wave
vector component for density fluctuation in four spatial dimensions. We
assume that the Levy exponent $\alpha $ for the Levy diffusion is
transferred into a tilt in the new $k$ direction. \ Since $E^{4}$ is
symmetric, the transfer effect must be distributed in all the four $k$
directions.

Second tilt change. In the further way, the loss of time cannot be absolute;
we need a time for the definition of expansion $\dot{R}(t)$. In the
speculation of section 5.1, this time is connected with photons as
''massless points'' from the filter elements. There remain only three $k$
directions which means a second change of the tilt, giving $1-n_{S}^{\prime }
$ for a density fluctuation before the inflation.\newline
\newline

Third tilt change. The next stage of the way is the hadronic inflation up to
the primordial stage from which the cosmic microwave background CMB is
measured today. The primordial tilt ($1-n_{S}$) follows after this third
change ($n_{S}^{\prime }\rightarrow n_{S}$) during the inflation.\newline
\newline

Assume that, via the above three changes, a small Levy tilt ($1-\alpha $) in
the original liquid results in another small but finite tilt in the
primordial state, measured as the deviation ($1-n_{S}$) from the Zeldovich
Harrison scaling $n_{S}=1$. Then we get from the three above priors, the
three changes, and equation (\ref{Eq.3.4}),

\begin{equation}
N_{G}=10^{(d\cdot \tilde{c}\cdot \alpha \varepsilon /2(1-c\alpha
)}=10^{3n_{S}\,\varepsilon /2(1-n_{S})}, \label{Eq.6.1}
\end{equation}%
where the three changes are collected by the factor $\tilde{c}$, $n_{S}=%
\tilde{c}\alpha $. I think that $\tilde{c}=0(1)$.\newline
\newline

Equation (\ref{Eq.6.1}) shows that a finite universe (containing a finite
number of galaxies $N_{G}<\infty $) as consequence of a Linde type expansion
can only exist for a finite tilt ($1-n_{S}>0$). For ($1-n_{S})\rightarrow 0$
we would have $N_{G}\rightarrow \infty $ and, because of equation (\ref%
{Eq.4.1}), the gravitational constant tends also to zero, $G_{N}\rightarrow
0 $. Without the finite tilt, i.e. for Zeldovich Harrison scaling of density
fluctuation, we would get an infinitely large universe ($N_{G}\rightarrow
\infty $) with no gravitation.\newline
\newline

The average CMB temperature fluctuation $\Delta T$ can be estimated from
Gauss statistics of the smallest representative subsystems, i.e. from the
number of galaxies, $N_{G}$. The consistency of Levy distribution for
estimation of the size of universe and the Gauss distribution for smaller
(and therefore faster) subsystems is explained in the text around figure 2.
The size of $\Delta T$ can therefore estimated by

\begin{equation}
\Delta T/T\approx 1/\sqrt{N_{G}}\approx 10^{-5}\text{ for
}N_{G}\approx 10^{10}. \label{Eq.6.2}
\end{equation}%
In the model, the finiteness of the CMB temperature fluctuation components $%
l(\Delta T_{l}>0)$ follows from the finiteness of the universe ($R<\infty $)
and is, from equation (\ref{Eq.6.1}), related to a finite tilt in the
density fluctuation.

\section*{7. Intrinsic flatness vs. gravitational dark energy}

The model allows to treat the cosmological constant $\Lambda $ exclusively
on the golden side of the Einstein equation; the dark energy can be reduced
to the flatness of the underlying $M_{4}$ tangent objects giving $\Omega
_{de}=\Omega _{\Lambda }=1-\Omega _{M}$. This degrades the dark energy
density parameter to $\Omega _{\Lambda }=\Omega _{\Lambda }(t)$ figures of
arithmetics of the golden side.\newline
\newline

The main idea is: Feeding the torus construction for the graviton (figure 3)
into the algebraic vdW spinor construction with physical coordinates for the
tangent object \cite{this part} defines the manifold diffeomorphism with $%
g_{ik}$ for the Einstein Hilbert action. This gravitational $g_{ik}$ is very
weak (10$^{-40}$) when compared with the electromagnetic nature of the flat $%
M_{4}$ tangent objects in a state with much vacuum. It is therefore the
geometric golden side of the Einstein equation which determines flatness,
not the energetic stony side.

\subsection*{7.1 Mach's principle of the model}

The filter elements mediate some kind of map between the universe and the
immediate environment on the $M_{4}$ tangent object of any isolated
culminating point particle (section 5.1). Because of conformal invariance
of the hidden space, the large filter elements pass over the whole universe
down to the small elements of size the particle ('s Compton wave length).
Assume that the central black holes of galaxies are mapped into spots on the
tangent objects, we get a {\it dollhouse universe} around any particle
(figure 7). These spots are related to the links of gravitons (${\Large %
\bullet }$ in figure 3) and can therefore be included in the Einstein
equation. Since the gravitation is a small interaction (equation \ref{Eq.4.1}%
), we find for isolated objects an additional shadow metric as
varying part of $g(x)$ over the flat $M_{4}$ tangent objects. This
defines a new, the Mach principle layer of the tangent objects
(\cite{this part}, appendix C there).

\begin{figure}[here]
\begin{center}
\includegraphics[width=0.25\linewidth]{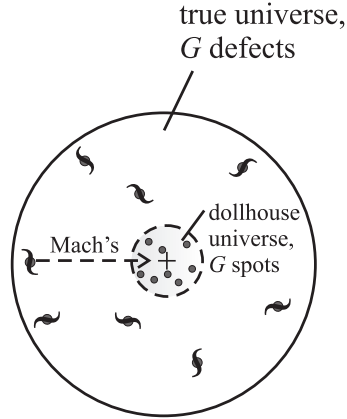}
\caption{Mach's principle maps (- - -\mbox{$>$}) the $G$ defects
(more precisely: the galaxies emerging from the $G$ defects of the
original or initial state) of the reality universe via the filter
elements into the $G$ spots on the (corresponding layer of the)
tangent objects around any culminating point particle: the
dollhouse universe. \label{fig7}}
\end{center}
\end{figure}

If this dollhouse universe from the filter is put on the stony side of the
Einstein equation, we would get some ''field'' $\phi $ that is neither
independent nor a quantizable Higgs field. It is located ''wherever'' a
particle is located, we do not find this field without a particle there. It
suits, however, to the equivalence principle of the gravitation theory,
since it represents the same situation around any particle.

\subsection*{7.2 Image constraints of the map: shadow metric $\longmapsto $
cosmological standard model}

From the viewpoint of a map from a finer model (original) to a coarser model
(image), the priors of the latter are image constraints of the map (\cite%
{this part}, section 4 there). The finer model may now be characterized by
\{flat $M_{4}$ tangent objects around isolated particles or matter
congregations, dollhouse universe, shadow metric\}, the coarser model by
\{total manifold with $g(x)$ metric, priors of cosmological standard
model\}. We consider the primordial stage (and later ones) of the universe:
{\it much vacuum}, i.e. only flat and very large $M_{4}$ tangent objects
around culminating-point constructions such as leptons, hadrons, dark matter
particles, baryons etc., and also around particle congregations such as
stars and black holes. We try to reduce the dark energy problem to flatness
of the underlying $M_{4}$ tangent objects in the vacuum of the universe.%
\newline
\newline

A prior of the general theory of relativity (gravitation theory) is the $%
g(x) $ field that is described by the Einstein Hilbert action,

\begin{equation}
S(\text{field})=(1/16\pi G_{N})\int (R_{{\small \square }}-2\Lambda )\sqrt{-g%
}d^{\,4}x, \label{Eq.7.1}
\end{equation}%
where $g=g(x)$ is then a Einstein Riemann metric assumed to be a manifold
for all things of the cosmos. $R_{{\small \square }}$ is the Ricci scalar
curvature. From equation (\ref{Eq.7.1}) follows by a variation $\delta (x)$
and by addition of a material term ($T_{ik}$) the Einstein equation. It
occurs in two forms, a geometric form with the cosmological parameter $%
\Lambda $ on the ''golden'' side, or an energetic form with $\Lambda $ on
the ''stony'' side:

\begin{equation}
\text{geometric, golden\quad }R_{ik}-(\frac{1}{2}R_{{\small \square }%
}+\Lambda )g_{ik}=(8\pi G_{N}/c^{4})T_{ik}, \label{Eq.7.2}
\end{equation}

\begin{equation}
\text{energetic, stony\quad }R_{ik}-\frac{1}{2}R_{{\small \square }%
}g_{ik}=(8\pi G_{N}/c_{4})T_{ik}+\Lambda g_{ik}. \label{Eq.7.3}
\end{equation}%
Flatness is a concept for $\Lambda $ on the golden side (curvature is
diminished by $\Lambda >0$ for given $T_{ik}$), dark energy is a concept of
the stony side (energy is raised by $\Lambda >0$). [In the literature, the
golden side is sometimes called a modification of gravity.] Absolutizing the
metric diffeomorphism, both sides cannot be distinguished. The $x/\psi $
separation \cite{this part}, however, can distinguish golden from stony, the
two equations are not longer equivalent.\newline
\newline

The additional prior of the conventional standard cosmology is the
Copernican principle (total homogeneity, isotropy, and cosmic time) that
usually is condensed in a Robertson Walker metric for the supposed manifold,

\begin{equation}
ds^{2}c^{2}dt^{2}-R^{2}(t)(\frac{dr^{2}}{1-kr^{2}}+r^{2}(d\theta
^{2}+\sin ^{2}\phi \,d\phi ^{2})), \label{Eq.7.4}
\end{equation}%
with $k=\{1,0,-1\}$ as indicator for the Gauss curvature of \{$S^{3}$, $%
E^{3} $, $L^{3}$\}, $L=$ Lobachevsky. At the moment, $k$ is a symmetry
parameter of a total Copernican solution, not of the Einstein equation.%
\newline
\newline

The total Copernican prior for the standard model of cosmology are therefore
the Friedmann Lamaitre (FL) equations,

\begin{equation}
\ddot{R}/R=\frac{1}{3}[\Lambda -4\pi G_{N}(\rho +3p/c^{2})],
\label{Eq.7.5}
\end{equation}

\begin{equation}
(\dot{R}/R)^{2}\equiv H^{2}=\frac{1}{3}(8\pi G_{N}\rho +\Lambda
)-kc^{2}/R^{2}, \label{Eq.7.6}
\end{equation}%
where ($\rho ,p$) are the energy density and the pressure of matter and
radiation. The expansion is accelerated by the dominance of $\Lambda $, with
the consequence that the Hubble rate $H$ increases with larger $\Lambda >0$.%
\newline
\newline

In the language of the stony side, an energy is connected with $\Lambda $,
the dark energy. For a time $t_{0}$ in the primordial or a later stage, e.g.
the expansion stage today (''why me, why now'' \cite{Turner}), the density
parameters for matter and radiation ($M$) and for $\Lambda $ are defined as

\begin{equation}
\Omega _{M}=\rho _{M}/\rho _{\text{crit}}\text{ , }\Omega
_{\Lambda }=\rho _{\Lambda }/\rho _{\text{crit}}\text{ , }\rho
_{\text{crit}}=3H_{0}/8\pi G_{N}, \label{Eq.7.7}
\end{equation}%
with $H_{0}=H(t_{0})$ and $\rho _{\Lambda }=\Lambda c^{3}/8\pi G_{N}$. Then
the space curvature $k$ is determined by the mysterious dark energy density $%
\rho _{\Lambda }$. For decreasing $\rho _{M}$ and given $\Omega _{\Lambda }$%
, the dominance of dark energy $\rho _{\Lambda }$ starts at a certain time $%
t_{0}^{\prime }$ (''now''); $\ddot{R}/R>0$ for large dominating $\rho
_{\Lambda }$ is also possible for a flat space.\newline
\newline

Experimental flatness means a local (not the total) property $k\approx 0$,
i.e. $\Omega _{M}+\Omega _{\Lambda }\approx 1$. The total $S^{3}$ or $L^{3}$
topology may be consistent with small local fluctuations near local
flatness, $\Omega _{M}+\Omega _{\Lambda }\approx 1$, especially for
sufficiently large $R$ and large $t$ (very much vacuum).\newline
\newline

In our language of the golden side, much vacuum between widely
isolated phenomena is necessary for a measurable influence of
$\Lambda $. In these large primordial scales, the vacuum is based
on the $M_{4}$ tangent objects from the biquaternion construction
\cite{this part}. They are flat because the hidden electromagnetic
structure is conformal. - Photons and gravitons are the only
massless (free) particles in the empty space. The ''cosmological''
photons in the shadow metric are not bound and change, therefore,
their frequency during the primordial expansion. The flat $M_{4}$
tangent objects do not come from photons alone but are bound to
the charges of our model (appendix C\cite{this part}). Their
$M_{4}$ properties, therefore, do not change during the expansion:
they remain exactly flat. The electromagnetic structure of the
tangent objects is always much ''stronger'' than the gravitons for
the shadow (equation (\ref{Eq.4.1})). This means that the empty
basic space (much vacuum) underlying the shadow metric is always
experimentally flat irrespective that the curved additional shadow
metric is decisive for the gravitational aspects of the primordial
universe. Both the flatness and the shadow curvature can be placed
in the FL equations, i.e. in the priors of conventional standard
cosmology.\newline
\newline

On the golden side, the above stony alternative is then to be interpreted by
the local relation $\Omega _{\Lambda }+\Omega _{M}\approx 1$. The stony term
$\rho _{\Lambda }=\Lambda c^{3}/8\pi G_{N}$ is not an energy in the
golden-side interpretation; i.e. the stony dark energy is geometrically
defined by golden flatness,

\begin{equation}
\Omega _{\Lambda }\approx 1-\Omega _{M}. \label{Eq.7.8}
\end{equation}%
The universe is experimentally flat since the primordial stage. The $\Omega
_{\Lambda }=\Omega _{\Lambda }(t)$ term in equation (\ref{Eq.7.8}) is
degraded to figures of arithmetics. There is no extra dark energy in the
cosmos and a search for it, e.g. in form of certain Higgs particles, is not
necessary from the viewpoint of our model.\newline
\newline

Guth's flatness is possible without GUTs.\newline
\newline
The destiny of our model universe is local flatness without matter, $\Omega
_{M}\rightarrow 0$, i.e. $\Omega _{\Lambda }\rightarrow 1$. If the universe
has got a total $S^{3}$ topology (section 5.1), then the universe in the
limit gets practically the state of the hidden charge, conformality without
gravitation, and the play can start anew, without any trace from the
foregoing universe, but with the same parameters so far as they are
determined by the culminating point filters. [On the other hand, this means,
e.g. that the new gravitational constant is not exactly the same, since
there is a variance of the number of $G$ defects inside the Fischer speckles
(of order of 10$^{-5}$ for $N_{G}=10^{10}$, cf. the end of section 5.3)].

\subsection*{7.3 Two properties from the dollhouse universe}

{\bf 1.} Equation of state parameter $w=-1$. Introduce a new parameter of
the stony side, $w=p/\rho $. Because of the huge number of vacuum elements
in our model, we have no fluctuation of the stony $\phi $ field. From
thermodynamics, especially from the fluctuation dissipation theorem FDT, we
get no response from no fluctuation. From the FL equations we obtain, via

\begin{equation}
\dot{\rho}=-3H\rho (1+\frac{p}{\rho }) \label{Eq.7.9}
\end{equation}%
the relevant response as

\begin{equation}
d\ln \rho /d\ln R=-3(1+w). \label{Eq.7.10}
\end{equation}%
From no response we have then $w=-1$ as announced. This justifies also the
only use of $\Lambda $ alone in section 7.2.\newline
\newline

{\bf 2.} The photon/baryon ratio is of order 10$^{10}$. The pass-over of the
large filter elements corresponds to one ''cosmic interaction diagram''
between a Feynman vertex of the baryon pro one galaxy spot. The spot acts as
catalyst, because in the large number approximation there is no feedback
from the Feynman vertex to the galaxy. The vertex forms a surviving baryon
in the model lepton capture reaction with a confinon and a prebaryon (\cite%
{this part}, equation (\ref{Eq.5.10}) there). The surviving rests on a
different partial chirality of the participating prebaryon (\cite{this part}%
, appendix C). Since the prebaryon has an electric charge, we assume that
there is about one photon pro one interaction. For $N_{G}=10^{10}$ we obtain
the announced ratio.

\section*{8. Conclusion}

The hidden charge model (shortly: the model) is introduced
by two concepts: \cite{this part} conceptual separation of mass and energy
behind $E=mc^{2}$, and \cite{second part} a new entity for an initial
cosmological liquid. The model remains in four dimensions, has a clearly
arranged particle spectrum confined by the TeV scale, and is supplied with a
rich thermodynamics. This allows a serious prosy representation and a
heuristic approach which two are suited to the early stage of the
conventional new standard cosmology. The model is much simpler and more
robust than highly sophisticated approaches, e.g. string theory.\newline
\newline

A hadronic inflation emerges because cold baryons can first be realized in
the length scale of $R\approx \lambda _{0}=1$fermi. For $R\ll \lambda _{0}$,
the filter elements for culminating-point baryons with $M_{4}$ tangent
objects are too small for quantum mechanics in a cold reality. Dense
dark-matter liquids, however could be realized for $R\ll \lambda _{0}$
because their tangent objects are $E^{4}$ Euclidean allowing points smaller
than $\lambda _{0}$. The sudden occurrence of baryons (and hadron pairs) at $%
R\approx \lambda _{0}$ generates a large (hadronic) amount of negative
pressure and therefore a violent pulse for inflationary expansion. During
its cosmic time scale the first small black holes emerge that allow a cold
Hawking unification of the four interactions.\newline
\newline

The new entities in the liquid are called Glarum Levy defects ($G$ defects).
It is finally the shaping power of the Levy distribution that forms the
galaxies of the universe and allows a relation between their number $N_{G}$
and the tilt ($1-n_{S}$) in the density fluctuation. $N_{G}\rightarrow
\infty $ would mean no tilt and no gravitation.\newline
\newline

The decrease of the filter element size during forming the culminating
points corresponds to a map of the universe in the $M_{4}$ tangent objects
near culminating point particles. In this dollhouse universe are, in
accordance with Mach's principle, the galaxies mapped to small spots that
are used to construct gravitons as charges from four hidden $\vartheta $
segments randomly connected by the spots. This defines gravitation, in the
vacuum as a shadow metric over flat $M_{4}$ tangent objects as basis. The
underlying flatness is based on electromagnetism, much stronger than the
weakness of the shadow metric from gravitation (factor 10$^{40}$). This
allows a purely geometric interpretation of the cosmological parameter $%
\Lambda $ (that was called dark energy before) as a consequence of this
flatness since the primordial stage of the universe, $\Omega _{\Lambda
}=1-\Omega _{M}$.\newline
\newline

Not all reality objects are quantized so as usually assumed. The electroweak
interaction remains quantized in the form that follows from Feynman path
integrals. Dark matter, if not isolated in the vacuum of Minkowski space, is
not conventionally quantized (e.g. in dense liquids). Somewhere between
these limiting cases is the quantization of strong interaction for hadrons,
in particular for baryons where $E^{4}$ tangent objects are inside from
prebaryons, and also the gravitation. Although gravitons are constructed as
charges, the possibility of a further general quantization of the Einstein
Riemann metric manifold must seriously be questioned. ''Dark energy'' as
geometrical flatness does not allow any quantization.\newline
\newline

Some conventional priors important for cosmology are substituted by the
model as follows: There are no Higgs particles because corresponding
eigensolutions are missing; the filter elements are not independent fields
and cannot be quantized because they do not fluctuate due to their huge
numbers. A hot big bang is not necessary, because cold dark matter particles
remain points also at large densities in the initial liquid for they cannot
be quantized in the Euclidean $E^{4}$ tangent objects there. The
cosmological parameter $\Lambda $ needs not to be an dark energy, because it
can be reduced to spatial flatness. $-$ In the frame of our hidden charge
model, a search for Higgs particles or independent dark energy is not useful.
\newline\newline

\section*{Note added at July 2010}

The above part of the paper corresponds to arXiv 0901.1050 v2 [physics.gen-ph] 20 Jan 2009, apart from twenty-two short adaptions and six corrections of mistakes. It seems useful to explain how this paradigmatic result was obtained.\newline\newline

I used a neural network method without computer. The author motivated systematically, since 1964, the neuron structure in his own head to find possibly new paradigmatic ideas in astroparticle physics. Partitions into a growing number of promising "remainders" of the known physics in growing specific time were used as changing starting conditions. At the end there are roughly twenty remainders. The iterations of the method were organized only by internal consistencies: completely \textit{inside} the paradigm, i.e. consistencies with all foregoing internal consistencies: between the changing remainders, between their binary relations, and the many new relations between foregoing consistencies.\newline\newline

Most of these internal iterations were documented by the author (all in German): the first with number 1 is of 4 May 1964 (physical point and its environment), the last with number 1381 is of 19 Oct 2003 ($\vartheta$ torus for graviton). Several hundreds of additional iterations became later implicit parts of rejected papers (about a dozen papers).\newline\newline

Two further examples. \textbf{1}. A new spin statistic theorem (SST) was firstly documented with number 352 of 16 May 1970 with headline SST. Its derivation from a threedimensional rotation excluded later SUSY (supersymmetry); see Part I, Sect. 4.1, Table I, and Eqs. (7), (8), and (9) there. \textbf{2}. The mass operator was indicated in number 899 of 8 Dec 1986 with headline pair generation, and was documented in number 948 of 9 Aug 1987 with the headline: Difference between charge (coordinates $\varphi_{1}$ and $\varphi_{2})$ and mass (coordinates $\vartheta$ and $\tau$), see again Part I, Section 4.1 there. The mystery of the Higgs particles and their connection with the flatness puzzle compensation in the universe (M. Veltmann, Facts and Mysteries in Elementary Particle Physics, World Scientific, New Jersey 2003, Section 10 there) gets a paradigmatic background from our application of the 90$^0$ polar $\vartheta$ coordinate for both, the mass operator and the graviton torus.\newline\newline

It turned out that the development of conventional new physics and the paradigmatic iterations diverge, i.e. they are drifting away one another, especially if the conventional (i.e. communicable) was organized as alternative ($y$/$n$) sequence. Communicable is defined by only a few (e.g. three or less) new concepts for understanding. An "off-site convergence" of the paradigmatic iterations was reached in 2008. This convergence was defined as the event where an important phenomenon not analyzed before could be explained without introduction of any new specified idea. This was the so-called "dark energy" of Part II (in Section 7.2 of our paper). As mentioned above, roughly two thousand internal paradigmatic iterations were necessary. Nevertheless, the paradigm did not become communicable and remains woolly today. The woolliness is assumed to decrease in the future.\newline\newline

To improve understanding, it seems useful to characterize the new paradigm by three "basic principles A, B, C". This will be tried here for the example of the 1-class solution (Part I, Tab. I there). The basic idea is a map from a hidden thing to the physical space (A $\overset{B}{\longmapsto }$C).\newline

\begin{description}
  \item[A.] Conformal hidden charge model (I, section 2.2, points (1), (2), (3) there).
  \item[B.] Existential instability as a consequence of the maximum for the lepton masses (I, Fig.6 there). This excludes relevant particles above the hadronic region (II, Eq.(12) here).
  \item[C.] Electroweak Feynman fabric in the physical space (I, Fig.2 there) as a consequence of the above A conformality.\newline
\end{description}

The relationship between the present new physics and an off-site converging paradigm was briefly presented in a Comment; see arXiv: 0906.2425 v2 [physics.gen-ph] 3 Aug 2009. This is a radical variant of a discussion in the common probability space for an alternative new physics sequence and the off-site development of a converging paradigm. Sociological aspects of the hierarchies in the present large collaborations are also presented. The radical excludes any perceptible probability to find a new paradigm inside the new physics.\newline\newline

In reality, the relationship may be gentler. From my experience with the common probability space, I think that the probability to find a new paradigm inside a sequence of communicable new physics remains rather small.\newline\newline

\section*{Acknowledgements}

This work was partly supported by the DFG Sonderforschungsbereich
SFB 418 and by the Fonds Chemische Industrie, FCI. The author
thanks in particular Steffen Trimper, the speaker of the SFB, and
has profited from stimulating discussions with E.W. Fischer
(Mainz), Dieter Richter (J\"{u}lich), Christoph Schick (Rostock),
Klaus Schr\"{o}ter (Halle), and our experimental group at the
Universit\"{a}t Halle collecting dynamic calorimetry data
supporting the belief to the characteristic length of glass
transition.

\end{document}